\documentclass[prl,aps,superscriptaddress,floatfix,tightenlines,twocolumn]{revtex4}
\usepackage{subfigure}
\usepackage{amsmath}
\usepackage{amssymb}
\usepackage{amsfonts}
\usepackage{epsfig}
\usepackage[dvips]{color}
\usepackage{bm}
\usepackage{times}

\begin{document}
\title{Zeno dynamics in quantum open systems}
\author{Yu-Ran Zhang}
\affiliation{Beijing National Laboratory for Condensed Matter Physics, Institute of Physics, Chinese Academy of Sciences, Beijing 100190, China}
\author{Heng Fan}
\email{hfan@iphy.ac.cn}
\affiliation{Beijing National Laboratory for Condensed Matter Physics, Institute of Physics, Chinese Academy of Sciences, Beijing 100190, China}
\affiliation{Collaborative Innovation Center of Quantum Matter, 100190 Beijing , China}
\date{\today}

\begin{abstract}
Quantum Zeno effect shows that frequent observations can slow down or even stop the unitary time
evolution of an unstable quantum system. This effect can also be regarded as a physical
consequence of the the statistical indistinguishability of neighboring quantum states.
The accessibility of quantum Zeno dynamics under unitary time evolution can be quantitatively
estimated by quantum Zeno time in terms of Fisher information. In this work, we
investigate the accessibility of quantum Zeno dynamics
in quantum open systems by calculating noisy Fisher information, in which a trace preserving
and completely positive map is assumed.
We firstly study the consequences of non-Markovian noise on quantum Zeno effect and give the exact forms
of the dissipative Fisher information and the quantum Zeno time. Then, for the
operator-sum representation, an achievable upper bound of the quantum Zeno time
is given with the help of the results in noisy quantum metrology.
It is of significance that the noise affecting the accuracy in the entanglement-enhanced
parameter estimation can conversely be favorable for the accessibility of quantum Zeno
dynamics of entangled states.
\end{abstract}

\maketitle

Quantum Zeno effect (QZE), coined as the Zeno's paradox in
quantum theory, states that an unstable quantum system, if
observed continuously, will never decay\cite{M&S}. Hence
we can slow down or even ``freeze'' the evolution of the
system by frequent measurements in its known initial state.
QZE is ascribed to two standard principles of quantum theory:
continuous unitary time evolution in the absence of
measurement and von Neumann projection postulate\cite{phyrep}.
The state of
the system need not remain frozen to its initial state, but
it could evolve in a multidimensional subspace, called ``Zeno subspace'', with
measurement projecting on this subspace\cite{F&P}.
QZE is anticipated to have significant applications in
protection of quantum states and creation of subspaces from decoherence
provided by a variety of sources, which are urgent for
robust quantum information processing\cite{enpr,CC,WXB,zoller,ye}.
There are also  experimental studies attempting at the confirmation of
QZE\cite{smerzi3} as well as its applications\cite{haroche,opt}.
Experiments on QZE have been performed mainly for oscillating systems\cite{exprlo2,exprlo1}, whilst there
are several attempts to observe QZE in truly decaying
states\cite{nature,exprl}.

QZE has become a focus of attention not only because it can be applied in robust
quantum information processing, but also because of its foundational implications
about the nature of quantum measurement\cite{phyrep} as well as indistinguishability
of state\cite{prd} and entanglement\cite{Smerzi2}. Recently, it has been shown that
Zeno dynamics can be comprehended as a physical consequence of the statistical
indistinguishability of neighboring quantum states in Hilbert space\cite{Smerzi}.
For example we consider a system Hamiltonian driving a pure state $e^{-iH_St}|\psi_0\rangle_S$,
and $m$ trials of projective measurements $M=|\psi_{0}\rangle_S\langle\psi_{0}|$ are
performed with equal time intervals $\tau=t/m$ during the dynamics. The survival probability to find
the system at its initial state can be written as $P(t)=1-(\tau/\tau_{Z})^2+\mathcal{O}(\tau^4)$
where $\tau_{Z}$ is the quantum Zeno time (ZT) in terms of Fisher information (FI) and
equals to the largest interval such that two states remain indistinguishable\cite{Smerzi}.
Thus, the accessibility of quantum Zeno dynamics can be quantitatively estimated by
 ZT that is obtained by calculating  FI. However, in real experiments there will always be some degree of
noise and limitation. Zeno dynamics of nonunitary physical process in quantum open system deserves further
investigation with fruitful results on the quantum Fisher information (QFI) in noisy
systems\cite{np,prl1}.

In quantum open systems, the dynamics of the system becomes ``noisy'' and nonunitary
due to interaction with an environment.  Generally, it can be described by a trace preserving and
completely positive (CP) map, named as a quantum channel.  Specifically, after the time unitary transformation
${U}_{SE}(t)$ acting on the state of system and environment $\rho_{SE}(0)$, we can obtain the reduced state of
system alone after a partial trace over the environment
${\rho}_{S}(t)=\mathcal{E}_{t}[{\rho}_{S}(0)]=\textrm{Tr}_{E}[{U}_{SE}(t)\rho_{SE}(0){U}^{\dag}_{SE}(t)]$\cite{N&C}.
When we assume that the system-environment state is initially decoupled $\rho_S(0)\otimes\rho_{E}(0)$,
the behavior of a quantum open system can be expressed by the operator-sum representation
$\mathcal{E}_{t}[{\rho}_{S}(0)]=\sum_{l}{\Pi}_{l}(t){\rho}_{S}(0){\Pi}^{\dag}_{l}(t)$ in terms of Karus operators.
Moreover, in many cases it turns out to be useful to formulate the dynamics of an open system by means of a quantum
Markovian master equation with Lindblad structure under the Born-Markovian approximation\cite{open}.
However, in many realistic physical systems the assumption of a Markovian dynamics relying on a number
of mostly rather drastic simplifications is not sufficient for modern applicaitons and non-Markovian dynamics
of an open system attracts nowadays increasing attention. Applying the time-convolutionless (TCL) projection
operator technique\cite{open}, one is able to obtain an exact master equation for the reduced system dynamics
in which the non-Markovianity are considered. 

In this work, we investigate the realizability of quantum Zeno dynamics in open system
via judging the indistinguishability of state with noisy FI.
We firstly investigate the consequences of non-Markovian noise on ZT via calculating noisy FI.
Two exactly solvable models are considered. 
Then, we study the quantum Zeno dynamics in an open system expressed by operator-sum representation\cite{N&C}.
In this case, we can utilize the general manifestation of quantum Zeno dynamics of unitary process proposed in Ref.~\cite{Smerzi}.
An achievable upper bound of the ZT is deduced via calculations of QFI using the variational methods in noisy
quantum metrology\cite{prl1}. Furthermore, it has been shown in Ref.~\cite{Smerzi} that the entangled state may have a
shorter ZT in unitary process than that of the separable state. We find that entangled state can have
a ZT with a similar scale as that of the separable state by interacting with the a suitable model of open system.
That is, the noise lowering the accuracy in the entanglement-enhanced parameter estimation can,
on the contrary, be beneficial to the accessibility of quantum Zeno dynamics of entangled states.

\section*{Results}
\noindent \textbf{Dissipative  Zeno dynamics via exact master equation.}
We consider an initial pure state ${\rho}_{S}(0)=|\psi_{0}\rangle_{S}\langle\psi_{0}|$
of a system $S$ evolving under the impact of noise. For simplicity, we assume
that the Hamiltionian for the system $H_{S}$ is time independent and
the dynamical equation describing the state is written in the interaction picture as $\dot{\rho}_{S}^{I}(t)=\mathcal{L}_{t}[\rho^{I}_{S}(t)]$ and
$\mathcal{L}_{t}[\circ]=-i[H_{\textrm{LS}}(t),\circ]+\sum_{k}\gamma_{k}(t)[A_{k}\circ A_{k}^{\dag}-\frac{1}{2}\{A_{k}^{\dag}A_{k},\circ\}]$.
As usual we set $\hbar=1$. $H_{\textrm{LS}}(t)=\sum_{k}S_{k}(t)A_{k}^{\dag}A_{k}$ is the Lamb shift Hamiltonian, and
$\{A_k\}$ is the set of Lindblad generators of the dynamical map. $S_{k}(t)$ is a time-dependent coefficient of the Lamb shift and $\gamma_{k}(t)$ denotes a time-dependent decay rate.
In Markovian evolutions, we have $\gamma_k(\tau) \geq 0$ $\forall k$ for $\tau\in[0,t]$,
while if any $\gamma_k(\tau)$ can be negative for some intervals, the dynamics of evolution will
be non-Markovian\cite{open}. Equivalently, the evolution in Schr\"{o}dinger picture can be expressed as
$\rho_{S}(t)=\mathcal{U}(t)[\rho_{S}(0)]=e^{-iH_St}\mathbb{T}e^{\int_{0}^tdt'\mathcal{L}_{t'}}[\rho_{S}(0)]e^{iH_St}$,
where $\mathbb{T}$ denotes time ordering.

We define the projective measurement applied in the quantum Zeno dynamics as $\mathcal{P}[\circ]=M\circ M$ with $M=|\psi_{0}\rangle_{S}\langle\psi_{0}|$.
A sequence of $m$ observations can repeatedly bring the system to the initial state with survival
probability $P(t)=\textrm{Tr}\left(\mathcal{V}(\tau)^m[\rho_S(0)]\right)$,
in which we define that $\mathcal{V}(\tau)=\mathcal{P}\mathcal{U}(\tau)\mathcal{P}$ and the interval is $\tau= t/m$.
For the case of small time intervals $\tau\ll t$ with a large enough number of trials $m \rightarrow\infty$, the
survival probability can be expanded in terms of intervals $\tau$ as
\begin{eqnarray}
P(t)=1-\frac{\mathcal{F}^{\textrm{d}}}{4m}t^{2}+\mathcal{O}(\tau^3)\simeq1-\left(\frac{\tau}{\tau^{\textrm{d}}_{Z}}\right)^{2}\label{eq1}
\end{eqnarray}
where
\begin{equation}
\mathcal{F}^{\textrm{d}}\equiv4(\Delta H_S)^2+2\sum_{k}\dot{\gamma}_{k}(0)\textrm{Cov}_{\rho_{S}(0)}(A_{k}^{\dag},A_{k})
\end{equation}
is called the dissipative Fisher information (d-FI)\cite{disspative} because
$\mathcal{F}^{\textrm{d}}\simeq {[\partial_{\tau}P(\tau)]^2}/\{P(\tau)[1-P(\tau)]\}$. Here,
$\textrm{Cov}_{\rho}(X,Y)\equiv\textrm{Tr}( XY\rho) - \textrm{Tr}( X\rho) \textrm{Tr}( Y\rho)$
is the covariance of observables $X$ and $Y$ with respect to the state $\rho$ and $(\Delta X)^2=\textrm{Cov}_{\rho}(X,X)$ denotes the variance.
The first term of $\mathcal{F}^{\textrm{d}}$ represents the contribution from
the system and the second term represents that from the dissipative bath. $\tau_{Z}^{\textrm{d}}=2/\sqrt{m\mathcal{F}^{\textrm{d}}}$ is called as the
dissipative quantum Zeno time (d-ZT) which coincides with the largest interval
such that the two states remain indistinguishable\cite{Smerzi2}. We can
conclude that the larger d-FI is, the shorter d-ZT will be and the harder quantum Zeno
dynamics is to be realized.

Then, we consider an exactly solvable model, the damped Jaynes-Cummings model (JCM)\cite{open},
to study the Zeno dynamics in non-Markovian environments.
A Hamiltonian of the total system is given by $H_{\textrm{tot}}=H_{S}+H_{B}+H_{I}$
where the system's Hamiltonian is $H_{S}=\omega_{0}\sigma_{+}\sigma_{-}$,
the Hamiltonian of vacuum reservoir is $H_{B}=\sum_{k}\omega_{k}b_{k}^{\dag}b_{k}$,
and $H_{I}=\sigma_{+}B+\sigma_{-}B^{\dag}$ denotes the interaction Hamiltonian
given that $B=\sum_{k}g_{k}b_{k}$ with $b_k$ ($b_k^\dag$) the boson annihilation (creation)
operator for the $k$th mode. Here, $\omega_0$ denotes the transition frequency
of the atom with ground state $|0\rangle$ and excited state states $|1\rangle$;
$\sigma_{x,y,z}$ are Pauli operators and $\sigma_{\pm}$ are the raising and lowering operators.
The initial state is given as $|\psi_0\rangle=(|0\rangle+|1\rangle)/\sqrt{2}$.
Given the Lorentzian spectral density
$J(\omega)={\lambda W^{2}}/\{{\pi[(\omega-\omega_{0})^{2}+\lambda^{2}]}\}$
with $W$ the transition strength and $\lambda$ the spectral width of the coupling,
we can obtain the master equation
$\dot{\rho}^{I}_{S}(t)=\gamma(t)[\sigma_{-}\rho_{S}^{I}(t)\sigma_{+}-\frac{1}{2}\{\sigma_{+}\sigma_{-},\rho_{S}^{I}(t)\}]$
where the time-dependent decay rate $\gamma(t)$ is written in two conditional forms:
\begin{eqnarray}
\gamma(t)=\left\{
\begin{array}{l l}
\frac{4W^{2}\sinh\left(\frac{dt}{2}\right)}{d\cosh\left(\frac{dt}{2}\right)+{\lambda}\sinh\left(\frac{dt}{2}\right)}&,\ W\leq{\lambda}/{2}\\
\frac{4W^{2}\sin\left(\frac{dt}{2}\right)}{d\cos\left(\frac{dt}{2}\right)+{\lambda}\sin\left(\frac{dt}{2}\right)}&,\ W>{\lambda}/{2}
\end{array}\right.
\end{eqnarray}
with $d=\sqrt{|\lambda^{2}-4W^{2}|}$.
In the weak coupling regime $W<\lambda/2$, $\gamma(t)$ is always positive which corresponds
to the Markovian process, while in the strong coupling regime $W\geq\lambda/2$, the function $\gamma(t)$
becomes negative within certain intervals of time, which displays the non-Markovianity\cite{QFIF}.
For both Markovian and non-Markovian regimes, we obtain the same results as $\dot{\gamma}(0)=4W^2$. The d-FI
is calculated as $\mathcal{F}^{\textrm{d}}=\omega_{0}^{2}+W^{2}$ which leads to
$\tau_{Z}^{\textrm{d}}=2/\sqrt{m(\omega_{0}^{2}+W^{2})}$. If there is no noise $W=0$,
the result will reduce to the unitary evolution case as discussed in Ref.~\cite{Smerzi2}. When the
transition strength becomes larger, d-FI grows and d-ZT decreases, which makes the Zeno
dynamics more difficult. Moreover, given a definite value of transition strength $W$, d-FI is independent of $\lambda$ and the result stays
unchanged for both Markovian  and non-Markovian noise. The interpretation of this extraordinary
result may be that in this example the initial dynamics characteristics of the open system do not
depend on  Markovianity or non-Markovianity.

\begin{figure}[t]
 \centering
\includegraphics[width=0.32\textwidth]{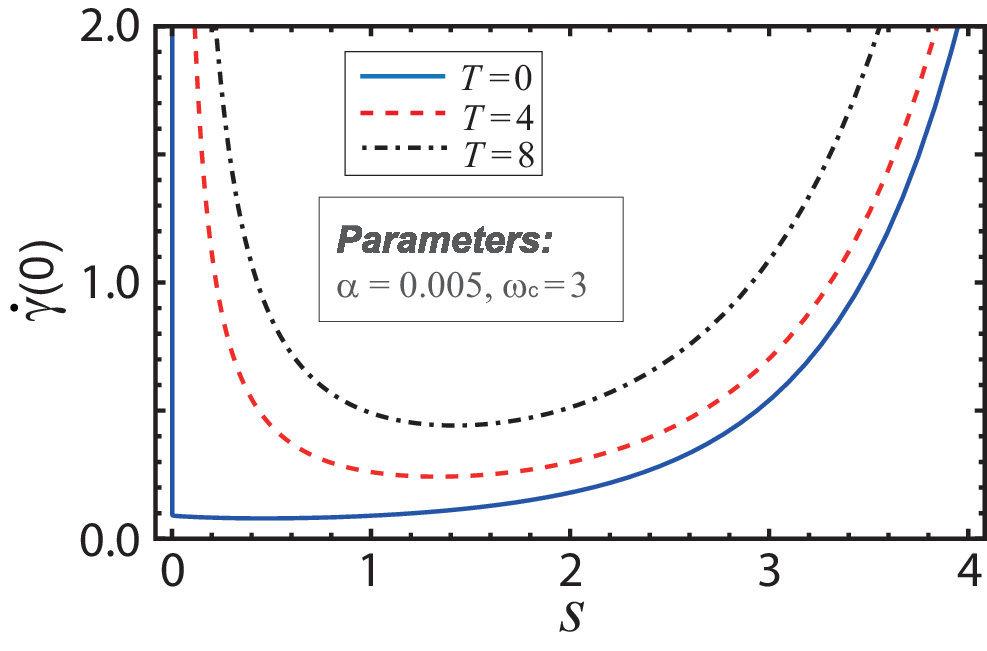}
\caption{\textbf{Figure 1 $|$ The first derivative of decay rate for $t=0$ agianst $s$.}
Parameters are set as $\alpha=0.01$ and $\omega_c=3$. Three cases
with three temperatures $T=0$, $4$ and $8$ are plotted by blue solid
line, red dashed line and black dot-dashed line, respectively.}\label{f2}
\end{figure}

Next, for the same initial state, we consider another exactly solvable model, the independent boson model, with
$H_{S}=\omega_{0}\sigma_{z}/2$, $H_{B}=\sum_{k}\omega_{k}b_{k}^{\dag}b_{k}$
and $H_{I}=\sum_{k}\sigma_{z}(g_{k}b_{k}^{\dag}+g_{k}^{*}b_{k})$. We consider the general
Ohmic-like spectral density with exponential cutoff
$J(\omega)=\alpha\omega_{c}^{1-s}\omega^{s}e^{-\omega/\omega_{c}}$: for $s=1$ it is Ohmic; for $s>1$ it is
super-Ohmic; for $s<1$ it is sub-Ohmic\cite{open}. The bath is assumed to be initially in a thermal
state: $\rho_{B}=\exp(-H_{B}/T)/\textrm{Tr}[{\exp(-H_{B}/T)}]$ given $T$ the temperature.
Then we can obtain the master equation as
$\dot{\rho}^{I}_{S}(t)=\gamma(t)\left[\sigma_{z}\rho_{S}^{I}(t)\sigma_{z}-\rho_{S}^{I}(t)\right]$
where
$\gamma(t)=2\int_{0}^{\infty}d\omega J(\omega)\coth\left(\frac{\omega}{2T}\right)\frac{\sin(\omega t)}{\omega}$.
For zero
temperature, the time-dependent decay rate can be carefully calculated as
$\gamma(t)=2\frac{\alpha\omega_{c}\Gamma(s)\sin[s\arctan(\omega_{c}t)]}{(1+\omega_{c}^{2}t^{2})^{s/2}}$
where $\Gamma(s)$ is the Euler Gamma function. If we consider non-zero temperature cases,
the first derivative of decay rate for $t=0$ can be exactly obtained as
\begin{eqnarray}
\dot{\gamma}(0)&=&2\alpha\omega_{c}^{1-s}T^{1+s}\Gamma(1+s)\times\nonumber\\&&\left[\zeta\left(1+s,1+\frac{T}{\omega_{c}}\right)+\zeta\left(1+s,\frac{T}{\omega_{c}}\right)\right]
\end{eqnarray}
where $\zeta(s,a)=\sum_{k=0}^{\infty}(k+a)^{-s}$ is the Hurwitz zeta function
(a generalized Riemann zeta function).
The d-FI and d-ZT may be calculated as $\mathcal{F}^{\textrm{d}}=\omega_{0}^{2}+2\dot{\gamma}(0)$ and
$\mathcal{\tau}_{Z}^{\textrm{d}}=2/\sqrt{m(\omega_{0}^{2}+2\dot{\gamma}(0))}$,
where  first derivative of decay rate for $t=0$ is shown in FIG.~1 given the parameters
$\alpha=0.005$ and $\omega_c=3$. It is shown that, as the temperature of bath becomes higher,
$\dot{\gamma}(0)$ increases, which makes the quantum Zeno dynamics more difficult. Moreover,
we find that for a definite temperature $T$, $\dot{\gamma}(0)$ declines first and increases then
as $s$ increases. Thus, to realize the Zeno dynamics for this model depends on the
temperature and the spectral density function of the bath.

\noindent \textbf{Quantum Zeno dynamics via operator-sum representation.}
A quantum process described in terms of an operator-sum representation is more general than
the one written down as a master equation \cite{N&C}.
For most circumstances, the noisy quantum channel can be written as a quantum dynamical map
$\mathcal{E}_{t}[\circ]$ in terms of Kraus operators $\{{\Pi}_{l}(t)\}$ with time $t$ the
parameter\cite{N&C}. The state evolves as
${\rho}_{S}(t)=\mathcal{E}_{t}[{\rho}_{S}(0)]=\sum_{l}{\Pi}_{l}(t){\rho}_{S}(0){\Pi}^{\dag}_{l}(t)$,
and the dynamical map is assumed to reduce to the identity map as $t=0$. This
nonunitary time evolution can also be transformed into a unitary time evolution operator
on an enlarged space $S+E$ for the system $S$ interacting with an environment $E$.
It can be expressed as
${\rho}_{S}(t)=\textrm{Tr}_{E}[{U}_{SE}(t)\rho_{SE}(0){U}^{\dag}_{SE}(t)]$,
where the initial state of $S+E$ is assumed to be  initially decoupled $\rho_{SE}(0)\equiv{\rho}_{S}(0)\otimes|0\rangle_{E}\langle0|$
and the unitary time evolution operator is assumed to have property
${U}_{SE}(t)|_{t\rightarrow0}=\mathbb{I}_{SE}$ with $\mathbb{I}_{SE}$ the identity
of enlarged space $S+E$. Therefore, we are able to use the results of QZE of unitary time evolution
discussed in Ref.~\cite{Smerzi} to investigate the quantum Zeno dynamics in open system.


After a sequence of $m$ observations using measurement operator
${M}=|\psi_{0}\rangle_{S}\langle\psi_{0}|\otimes\mathbb{I}_{E}$ with $\mathbb{I}_{E}$
the identity of environment $E$, the state of the system $S$ stays unchange with a survival
probability
$P(t)=\textrm{Tr}[{V}(\tau)^{m}\rho_{SE}(0){V}^{\dag}(\tau)^{m}]$,
where ${V}(\tau)={M}{U}_{SE}(\tau){M}$ and still the interval is $\tau=t/m$.
For infinitesimal time intervals $\tau\ll t$ with a large enough number of trials
$m\rightarrow\infty$, the survival probability can be expanded in terms of
intervals $\tau$ (See Methods for details.)
\begin{eqnarray}
P(t)=1-\frac{\mathcal{C}_{Q}(\tau,\mathcal{H})}{m}t^{2}+\mathcal{O}(\tau^3)
\label{prob}
\end{eqnarray}
in which the Hermitian operator ${\mathcal{H}}\equiv{H}_{SE}-{M}{H}_{SE}{M}$ and
the Hermitian generator of displacement in parameter $t$ is\cite{Generalized}
${H}_{SE}
\equiv-i[{d{U}_{SE}(t)}/{dt}]_{t\rightarrow0}$. Here, $\mathcal{C}_{Q}(\tau,\mathcal{H})=\Delta\mathcal{H}^2$
can be regarded as the QFI for the Hermitian generator $\mathcal{H}$ of the enlarged system $S+E$.
The information about the interval $\tau$ when system $S$ and environment
$E$ are monitored together is larger or equal to that obtained when merely the system
$S$ is monitored. Therefore, the QFI of the enlarged system $S+E$ gives an achievable upper bound
of the QFI of system $S$\cite{np,prl1}: $\mathcal{F}_{Q}(\tau)\leq\mathcal{C}_{Q}(\tau,{\mathcal{H}})$.
The ZT of the enlarged system $S+E$ has a time scale that is upper bounded
by the smallest path interval of QFI of system alone such that two states are
statistically distinguishable\cite{prd,Smerzi}
\begin{eqnarray}
\tau_{Z}=\frac{2}{\sqrt{m\mathcal{C}_{Q}(\tau,{\mathcal{H}})}}\leq\frac{2}{\sqrt{m\mathcal{F}_{Q}(\tau)}}=\tau_{m}
\label{main1}
\end{eqnarray}
where $\tau_{m}$ is the largest interval for QZE of the noisy quantum channel.
The achievable maximum of $\tau_{Z}$ is obtained when
$\mathcal{C}_{Q}(\tau,{\mathcal{H}})$ reaches its minimum, which is tantamount
to calculating the QFI of a noisy quantum channel $\mathcal{F}_{Q}(\tau)$  corresponding
to the entire unitary time evolution $\tilde{U}_{SE}(t)=\exp(-i{\mathcal{H}}t)$\cite{prl1}.

The QFI of a noisy quantum channel can be achieved over all the possible
and effective operator ${h}_{E}(\tau)\equiv i\frac{d{u}_{E}^{\dag}(\tau)}{d\tau}{u}_{E}(\tau)$
with unitary operator ${u}_{E}(\tau)$ acting solely on the space of environment $E$\cite{prl1,N&C}
(see Methods for details).
Therefore, we can obtain \cite{prl1}
\begin{eqnarray}
\mathcal{F}_{Q}(\tau)=\min_{\{{h}_{E}(\tau)\}}\mathcal{C}_{Q}(\tau,\tilde{\mathcal{H}}(\tau))
\label{main2}
\end{eqnarray}
where we define the Hermitian operator as $\tilde{\mathcal{H}}(\tau)\equiv{\mathcal{H}}+\tilde{h}(\tau)$
and
$\tilde{h}(\tau)\equiv\tilde{U}_{SE}^{\dag}(\tau){h}_{E}(\tau)\tilde{U}_{SE}(\tau)$.
The unitary time evolution ${u}_{E}(t){U}_{SE}(t)$ of environment $E$ together
with system $S$ does not lead to more information about parameter $\tau$ than
that obtained by system $S$ itself. We are also able to find a set of equations for the
optimum effective Hermitian operator ${h}_{E}^{\textrm{opt}}$ that minimizes
$C_{Q}(\tau,\tilde{\mathcal{H}}(\tau))$.

There are, in fact, infinite different unitary evolutions of the enlarged system
$S+E$ corresponding to the same operator-sum representation of system $S$, since it has the unitary freedom ${u}_{E}(t)$\cite{N&C,nc}. Each
one gives a different value of QFI $C_{Q}(\tau,\tilde{\mathcal{H}}(\tau))$. Even so,
the maximum ZT $\tau_{m}$ leads to an interesting and important physical insight:
there is always an environment $E$ making the quantum Zeno dynamics most
accessible. This result is promising to protect quantum information from decoherence,
especially for entangled states.

\noindent \textbf{$N$-qubit quantum Zeno dynamics in quantum open systems.}
Different states, entangled or separable, with different values of QFI lead to different
ZT scales\cite{Smerzi}. QFI of a separable state of an $N$-qubit system governed by a local
Hamiltonian ${H}_{S}=\omega_{0}\sum_{i=1}^{N}\bm{\sigma}^{i}\cdot\bm{n}^{i}$,
is bounded by $\mathcal{F}_{Q}\leq{N}\omega_{0}^{2}$, where
$\bm{\sigma}^{i}=(\sigma_x^i,\sigma_y^i,\sigma_z^i)$ is the vector of three Pauli matrices
acting on the $i$-th qubit and $\bm{n}^{i}$ is a unit vector. Therefore,
$\mathcal{F}_{Q}>N\omega_{0}^{2}$ is a sufficient condition for the presence of entanglement
\cite{Smerzi2}. As a consequence of larger QFI, the quantum Zeno dynamics of entangled states
may require a much higher rate of projective measurements than that of separable
ones as the number of qubits grows too large\cite{Smerzi}. Next, we will exemplify that the
quantum Zeno dynamics of maximal entangled states may only require a similar number
of measurements as that of separable states if the system interacts with a proper environment.

We consider an $N$-qubit system of which each qubit merely interacts
with a corresponding environmental qubit. It 
can be described as a unitary operator onto an enlarged system $S+E$ by
tracing out all the environmental qubits:
\begin{eqnarray}
{U}_{SE}(t)=\prod_{i=1}^{N}e^{-i\omega_{0}{\sigma}^{i}_{z,S}t/2}e^{-i\Gamma{\sigma}^{i}_{z,S}{\sigma}^{i}_{x,E}t/2},
\label{1q}
\end{eqnarray}
where ${\sigma}^{i}_{z,S}$ is the Pauli matrix acting on the $i$th system qubit
and ${\sigma}^{i}_{x,E}$ is on its environment qubit. The initial state of the
environment qubits is set as $|0\rangle^{\otimes N}_{E}$.
Given the system's initial state a maximal entangled state
$(|0\rangle_{S}^{\otimes{N}}+|1\rangle_{S}^{\otimes{N}})/\sqrt{2}$,
we can obtain the QFI as (See Methods.):
\begin{eqnarray}
\mathcal{F}^{\textrm{en}}_{Q}(\tau)=\frac{\omega_{0}^{2}N^{2}}{N\tan^{2}(\Gamma\tau)+1}+N\Gamma^{2},
\label{conclu2}
\end{eqnarray}
which has a limit $\mathcal{F}^{\textrm{en}}_{Q}(\tau)\rightarrow{N}[\omega_{0}^{2}\cot^{2}(\Gamma\tau)+\Gamma^{2}]$
as $N\rightarrow\infty$. It leads to the upper bound of ZT $\tau_{Z}^{\textrm{en}}\leq\tau_{m}^{\textrm{en}}=2/\sqrt{m}\mathcal{F}^{\textrm{en}}_{Q}(\tau)$.
The QFI of a separable initial state of the system with form
$[(|0\rangle_{S}+|1\rangle_{S})/\sqrt{2}]^{\otimes{N}}$  for the same quantum dynamical map is
\begin{eqnarray}
\mathcal{F}^{\textrm{se}}_{Q}(\tau)=N[\omega_{0}^{2}\cos^{2}(\Gamma\tau)+\Gamma^{2}],
\label{conclu3}
\end{eqnarray}
and $\tau_{Z}^{\textrm{se}}\leq\tau_{m}^{\textrm{se}}=2/\sqrt{m}\mathcal{F}^{\textrm{se}}_{Q}(\tau)$.
We can conclude that  the ratio
$\mathcal{F}^{\textrm{en}}_{Q}(\tau)/\mathcal{F}^{\textrm{se}}_{Q}(\tau)$ is independent of $N$
for an infinitely great $N$, no matter how small the interaction between system and environment is.
This result conforms to the conclusions of entanglement-enhanced parameter estimation in open
systems: the use of maximal entangled states fails to provide higher resolution as compared to
using separable states where decoherence exists\cite{MP1,MP3,MP4,pe1,pe2}.
\begin{figure}[t]
 \centering
\includegraphics[width=0.35\textwidth]{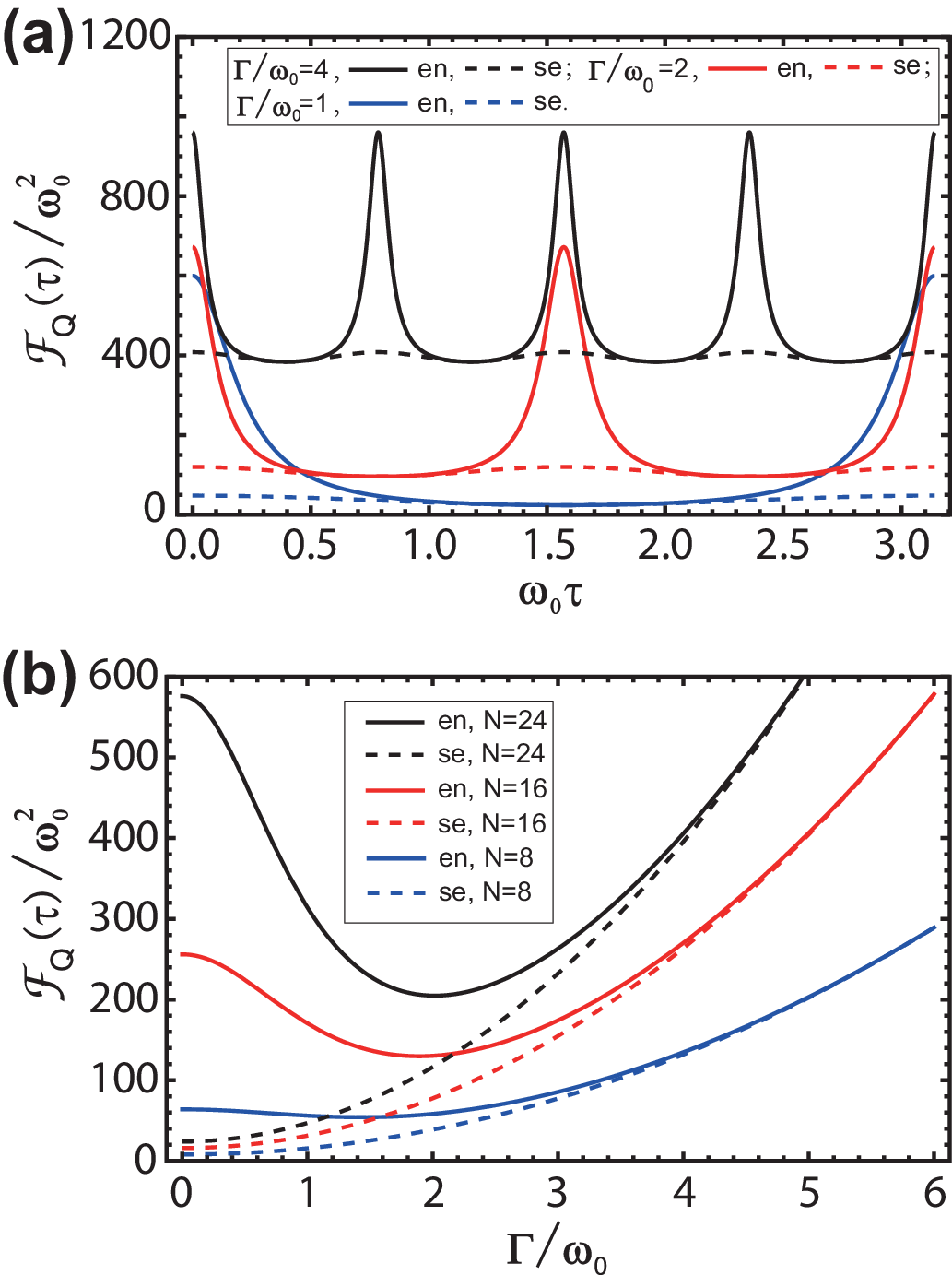}
\caption{\textbf{Figure 2 $|$ QFI of separable state and entangled state.} Solid lines are for QFI of
entangled state (en) and dashed lines are for separable state (se). ({a})
$\mathcal{F}_{Q}(\tau)/\omega_{0}^2$ against $\omega_{0}\tau$ with different
interaction strength: $\Gamma/\omega_{0}=1$ (blue lines), $2$ (red lines) and $4$
(black lines). The qubit number is set as $N=24$.
({b}) $\mathcal{F}_{Q}(\tau)/\omega_{0}^2$ against  $\Gamma/\omega_{0}$ with
different qubit numbers: $N=8$ (blue lines), $16$ (red lines) and $24$
(black lines). The time interval is set as  $\omega_{0}\tau=0.2$.}\label{f3}
\end{figure}

Specifically, we can obviously see from FIG.~2(a) that
$\mathcal{F}^{\textrm{en}}_{Q}(\tau)\simeq\mathcal{F}^{\textrm{se}}_{Q}(\tau)$
for some time interval. In FIG.~2(b) we find that when the strength of environment
$\Gamma$ is weak, $\mathcal{F}^{\textrm{en}}_{Q}(\tau)$ is larger than $\mathcal{F}^{\textrm{se}}_{Q}(\tau)$,
i.e., quantum Zeno dynamics of entangled states may be harder to realize. As
the increase of $\Gamma$, quantum Zeno dynamics of both cases are equally accessible.
However, given ``strong environment'' $\Gamma/\omega_0\ll1$,  both $\mathcal{F}^{\textrm{en}}_{Q}(\tau)$ and $\mathcal{F}^{\textrm{se}}_{Q}(\tau)$ tend to be infinity and the ZT
is confined to be so small that it makes the quantum Zeno dynamics nearly
accessible as predicted in Ref.~\cite{Smerzi}. It is thus significant that the appropriate
environmental interaction can be favourable for realizing QZE of entangled states
compared with the case of unitary time evolution in closed system.
This effect may be explained by the fact that some decoherence acts like an effective
further continuous measurement on the system, therefore making the QZE more accessible.
Besides, we can also figure out the optimal model of  environment given the definite form
of states and the definite noisy channel, which is shown in Methods.  Our theory is also able
to settle the case for states which are not maximally entangled but may bring new
interesting results of QZE in open system.

\section*{Discussion}
We have investigated the accessibility of quantum Zeno dynamics in quantum open systems. The
quantum Zeno dynamics in non-Markovian noise has been studied with d-FI and d-ZT when the exact
master equations are used to describe the quantum open system.  The more general description
using operator-sum representation of the open system has also been considered and investigated.
Due to the external unitary freedom of this description, an achievable upper bound of
ZT is deduced via the variational methods. Although entanglement will enhance the speed of evolution
and hinder QZE for unitary process\cite{MP2}, we have been examplified that the quantum Zeno dynamics
of maximal entangled states can be realized much easier when they interacts with the proper
environment than without noise.  That is, the noise affecting the accuracy in the quantum parameter estimation can
conversely be favorable for the accessibility of quantum Zeno dynamics of entangled states.
Our work will help to stablize the system of entangled states against time evolution and noise
in many quantum systems\cite{pe2,exprlo2,zoller,smerzi3}.

\section*{Methods}
\noindent \textbf{QFI and Zeno dynamics.}
Given a unitary dynamics $e^{-iHt}|\psi_{0}\rangle$ and $m$ trials of projections
$M=|\psi_{0}\rangle\langle\psi_0|=\rho_0$, the survival probability of the Zeno dynamics is
$P(t)=\textrm{Tr}[V(\tau)^m\rho_0V^{\dag}(\tau)^m]$ where $\tau=t/m$ and $V(\tau)\equiv Me^{-iH\tau}M$.
The survival probability  for small time intervals $\tau$ can be expanded as\cite{Smerzi}
$P(t)=1-m\Delta^2\mathcal{H}\tau^2+\mathcal{O}(\tau^3)$ where $\mathcal{H}=H-MHM$. The leading role in the
theory of this work is played by the FI\cite{SLD}:
$\mathcal{F}(\tau)\equiv\sum_{\xi}{P(\xi|\tau)}\left[{{\partial}_{\tau} \ln P(\xi|\tau)}\right]^{2}$
where $P(\xi|\tau)=\textrm{Tr}[\rho(\tau)E(\xi)]$ given $\{E(\xi)\}$ a set of POVMs.
QFI is obtained by exploiting the maximum of FI among all the
possible POVMs, and for unitary evolution $e^{-i\mathcal{H}t}|\psi_{0}\rangle$, it can
be expressed in a simple analytical expression\cite{seth}:
$\mathcal{F}_{Q}(\tau)=4\Delta^{2}\mathcal{H}$ when we take $t$ as the unknown parameter.
Therefore, the Zeno dynamics for unitary evolution with generator $H$ can be approximately
described by the QFI of the unitary evolution with Hermitian generator $\mathcal{H}$.

Generally, if the survival probability for intevals, $\tau\ll1$, can be expanded as
$P(\tau)=1-F\tau^2/4+\mathcal{O}(\tau^3)$, we can calculate the FI as
$\mathcal{F}=[\partial_{\tau}P(\tau)]^2/\{P(\tau)[1-P(\tau)]\}\simeq F$.
For the master equation approach, the survival probability
$P(t)=\textrm{Tr}\left(\mathcal{V}(\tau)^m[\rho_S(0)]\right)={}_S\langle\psi_0|\rho_S(\tau)|\psi_0\rangle_S^m$
and we can expand the density operator  $\rho_{S}(\tau)=\rho_{S}(0)+\dot{\rho}_{S}(0)\tau+\ddot{\rho}_{S}(0)\frac{\tau^{2}}{2}+\mathcal{O}(\tau^{3})$
for small time intervals. We can obtain the first order derivative as
$\dot{\rho}_{S}(0)=-i[H_{S},\rho_{S}(0)]+\dot{\rho}_{S}^{I}(0)$
and for most physical cases without the Markovian approximation,
$\gamma_{k}(0)=S_{k}(0)=0$ and $\dot{\rho}^{I}_{S}(0)=0$ hold for all spectral densities
\cite{open,MP1,anti} with which we obtain $\langle\psi|\dot{\rho}_{S}(0)|\psi\rangle=0$.
For the second order derivative, we have
$\langle\psi|\ddot{\rho}_{S}(0)|\psi\rangle=-2\Delta H^2_{S}-\sum_{k}\dot{\gamma}_{k}(0)\textrm{Cov}_{\rho_{S}(0)}(A_{k}^{\dag},A_{k})$,
with which Eq.~(\ref{eq1}) can be proved.

\noindent \textbf{Noisy QFI of  maximal entangled state of $N$ qubits.}
In the noisy model expressed in Eq.~(\ref{1q}), the Hermitian operator may be calculated as
${\mathcal{H}}=\sum_{i=0}^{N}\left({\omega_{0}}{{Z}}_{S}^{i}+{\Gamma}{Z}^{i}_{S}{X}^{i}_{E}\right)/2$
such that ${U}_{SE}(t)=\tilde{U}_{SE}(t)$. In accordance with the symmetry of maximal
entangled states, the general form of the Hermitian operator acting solely on
environment $E$ may be expressed as\cite{prl2}
\begin{eqnarray}
{h}_{E}=\sum_{i=1}^{N}[\alpha(\tau){\sigma}_{x,E}^{i}+\beta(\tau)\sigma_{y,E}^{i}+\gamma(\tau){\sigma}_{z,E}^{i}].
\end{eqnarray}
where $\alpha(\tau)$, $\beta(\tau)$ and $\gamma(\tau)$ are variables in terms of
parameter $\tau$. We can calculate the exact form of $\Delta^{2}\tilde{\mathcal{H}}$ as
\begin{eqnarray}
\Delta^{2}\tilde{\mathcal{H}}&=&(N^{2}\omega_{0}^{2}+N\Gamma^{2})/4\nonumber\\
&+&N[\alpha(\tau)^{2}+\beta(\tau)^{2}\cos^{2}(\Gamma\tau)+\gamma(\tau)^{2}\sin^{2}(\Gamma\tau)]\nonumber\\
&+&N^{2}\beta(\tau)^{2}\sin^{2}(\Gamma\tau)-\omega_{0}N^{2}\sin(\Gamma\tau)\beta(\tau).
\end{eqnarray}
Then, we minimize
$\mathcal{C}_{Q}(\tau,\mathcal{H})=\Delta^{2}\tilde{\mathcal{H}}$
over $\alpha(\tau)$, $\beta(\tau)$ and $\gamma(\tau)$ for any
value of $\tau$ with conditions
$\partial_{\alpha}\Delta^{2}\tilde{\mathcal{H}}=\partial_{\beta}\Delta^{2}\tilde{\mathcal{H}}=\partial_{\gamma}\Delta^{2}\tilde{\mathcal{H}}=0$.
The optimal parametric equations may be obtained as
$\alpha^{\textrm{opt}}(\tau)=\gamma^{\textrm{opt}}(\tau)=0$ and
$\beta^{\textrm{opt}}(\tau)=\frac{\omega_{0}N\sin(\Gamma\tau)}{2[N\sin^{2}(\Gamma\tau)+\cos^{2}(\Gamma\tau)]}$.
Thus, the QFI of the noisy system $\mathcal{F}_{Q}^{\textrm{en}}(\tau)=\min{4\Delta^{2}\tilde{\mathcal{H}}(\tau)}$
is obtained as shown in Eq.~(\ref{conclu2}). For the separable state, we let $N=1$ and obtain the QFI
using the additivity of FI.

With the optimal parametric equations, we are still able to obtain the exact form of the optimal
environment that maximizes the ZT. With the optimal Hermitian operator $h_{E}^{\textrm{opt}}(\tau)=\sum_{i=1}^{N}\beta^{\textrm{opt}}(\tau)\sigma_{y,E}^{i}$,
the optimal unitary time evolution is written as
${U}^{\textrm{opt}}_{SE}={u}^{\textrm{opt}}_{E}{U}_{SE}$
with ${u}^{\textrm{opt}}_{E}=\exp[{-i\int_0^\tau{h}^{\textrm{opt}}(s)ds}]$.

\section{Acknowledgments}
We would like to thank Augusto Smerzi, Wei-Bin Yan and Ying-Jie Zhang for useful discussions.
This work was supported by the ``973'' Program (2010CB922904),
NSFC (11175248)
grants from the Chinese Academy of Sciences.

\section{Author contributions}
Y.-R. Z. and H.F. proposed the model. Y.-R.Z. calculates the results. Y.-R.Z.and H.F.
analyzed the results. Y.-R.Z. and H.F. wrote the paper.

\section{Competing financial interests}
The authors declare no competing financial interests.



\begin{thebibliography}{99}

\bibitem{M&S} Misra, B. \& Sudarshan, E. C. G.
The Zeno's paradox in quantum theory.
\emph{J. Math. Phys.} \textbf{18}, 756 (1997).

\bibitem{phyrep} Koshino, K. \& Shimizu, A.
Quantum Zeno effect by general measurements.
\emph{Phys. Rep.} \textbf{412}, 191 (2005).

\bibitem{F&P} Facchi, P. \& Pascazio, S.
Quantum Zeno subspaces.
\emph{Phys. Rev. Lett.} \textbf{89}, 080401 (2002).

\bibitem{CC} Paz-Silva, G. A., Rezakhani, A. T., Dominy, J. M. \& Lidar, D. A.
Zeno effect for quantum computation and control.
\emph{Phys. Rev. Lett.} \textbf{108}, 080501 (2012).

\bibitem{enpr} Maniscalco, S., Francica, F., Zaffino, R. L., Gullo, N. L. \& Plastina, F.
Protecting entanglement via the quantum Zeno effect.
\emph{Phys. Rev. Lett.} \textbf{100}, 090503 (2008).

\bibitem{WXB} Wang, S. C., Li Y., Wang, X. B. \& Kwek, L. C.
Operator quantum Zeno effect: protecting quantum information with noisy two-qubit interactions.
\emph{Phys. Rev. Lett.} \textbf{110}, 100505 (2013).

\bibitem{zoller} Stannigel, K., Hauke, P., Marcos, D., Hafezi, M., Diehl, S., Dalmonte, M. \& Zoller, P.
Constrained dynamics via the Zeno effect in quantum simulation: implementing non-Abelian lattice gauge theories with cold atoms.
\emph{Phys. Rev. Lett.} \textbf{112}, 120406 (2014).

\bibitem{ye} Zhu, B. \emph{et al.}
Suppressing the loss of ultracold molecules via the continuous quantum Zeno effect.
\emph{Phys. Rev. Lett.} \textbf{112}, 070404 (2014).

\bibitem{smerzi3} Schafer, F., Herrera, I., Cherukattil, S., Lovecchio, C., Cataliotti, F. S., Caruso, F. \& Smerzi, A.
Experimental realization of quantum zeno dynamics.
\emph{Nature Commnun.} \textbf{5}, 3194 (2014).

\bibitem{haroche} Signoles, A., Facon, A., Grosso, D., Dotsenko, I., Haroche, S., Raimond, J. M., Brune, M. \& Gleyzes, S.
Confined quantum Zeno dynamics of a watched atomic arrow.
\emph{Nature Phys.} \textbf{10}, 715 (2014).

\bibitem{opt} McCusker, K. T., Huang, Y. P.,  Kowligy, A. S. \& Kumar, P.
Experimental demonstration of interaction-free all-optical switching via the quantum Zeno effect.
\emph{Phys. Rev. Lett.} \textbf{110}, 240403 (2013).


\bibitem{exprlo2} Itano, W. M., Heinzen, D. J., Bollinger, J. J. \& Wineland, D. J.
Quantum Zeno effect.
\emph{Phys. Rev. A} \textbf{41}, 2295 (1990).

\bibitem{exprlo1} Bernu, J., Del\'{e}glise, S., Sayrin, C., Kuhr, S.,  Dotsenko, I.,  Brune, M.,  Raimond, J. M. \&  Haroche, S.
Freezing coherent field growth in a cavity by the quantum Zeno effect.
\emph{Phys. Rev. Lett.} \textbf{101}, 180402 (2008).

\bibitem{nature} Wilkinson, S. R., Bharucha, C. F., Fischer, M. C., Madison, K. W., Morrow, P. R., Niu, Q., Sundaram, B. \& Raizen, M. G.
Experimental evidence for non-exponential decay in quantum tunnelling.
\emph{Nature} \textbf{387}, 575 (1997).

\bibitem{exprl} Fischer, M. C., Gutierrez-Medina, B. \& Raizen, M. G.
Observation of the quantum Zeno and anti-Zeno effects in an unstable system.
\emph{Phys. Rev. Lett.} \textbf{87}, 040402 (2001).

\bibitem{prd} Wootters, W. K.
Statistical distance and Hilbert space.
\emph{Phys. Rev. D} \textbf{23}, 357 (1981).

\bibitem{Smerzi2} Pezz\'{e}, L. \& Smerzi, A.
Entanglement, nonlinear dynamics, and the Heisenberg limit.
\emph{Phys. Rev. Lett.} \textbf{102}, 100401 (2009).

\bibitem{Smerzi} Smerzi, A.
Zeno dynamics, indistinguishability of state, and entanglement.
\emph{Phys. Rev. Lett.} \textbf{ 109}, 150410 (2012).

\bibitem{np} Escher, B. M., de Matos Filho, R. L. \& Davidovich, L.
General framework for estimating the ultimate precision limit in noisy quantum-enhanced metrology.
\emph{Nature Phys.} \textbf{7}, 406 (2011).

\bibitem{prl1} Escher, B. M.,  Davidovich, L.,  Zagury, N. \&  de Matos Filho, R. L.
Quantum metrological limits via a variational approach.
\emph{Phys. Rev. Lett.} \textbf{109}, 190404 (2012).

\bibitem{N&C} Nelsen, A. S. \& Chuang, I. L.
\emph{Quantum Computation and Quantum Information}
(Cambridge University Press, Cambridge, England, 2000).

\bibitem{open} Breuer, H.-P. \& Petruccione, F.
\emph{The Theory of Open Quantum System}
(Oxford University Press, New York, United States, 2003).

\bibitem{disspative} Alipour, S., Mehboudi, M. \& Rezakhani, A. T.
Quantum metrology in open systems: dissipative Cram\'{e}r-Rao bound.
\emph{Phys. Rev. Lett.} \textbf{112}, 120405 (2014).

\bibitem{QFIF} Lu, X. M., Wang, X. G. \& Sun, C. P.
Quantum Fisher information flow and non-Markovian processes of open systems.
\emph{Phys. Rev. A} \textbf{82}, 042103 (2010).

\bibitem{Generalized} Boixo, S., Flammia, S. T., Caves, C. M. \& Geremia, J. M.
Generalized limits for single-parameter quantum estimation.
\emph{Phys. Rev. Lett.} \textbf{98}, 090401 (2007).

\bibitem{nc} Demkowicz-Dobrza\'{n}ski, R., Ko{\l}ody\'{n}ski, J. \&  Gu\c{t}\v{a}, M.
The elusive Heisenberg limit in quantumenhanced metrology.
\emph{Nature Commun.} \textbf{3}, 1063 (2012).

\bibitem{MP1} Chin, A. W., Huelga, S. F. \& Plenio, M. B.
Quantum Metrology in Non-Markovian Environments.
\emph{Phys. Rev. Lett.} \textbf{109}, 233601 (2012).

\bibitem{MP3} Huelga, S. F., Macchiavello, C., Pellizzari, T., Ekert, A. K., Plenio, M. B. \& Cirac, J. I.
Improvement of frequency standards with quantum entanglement.
\emph{Phys. Rev. Lett.} \textbf{79}, 3865 (1997).

\bibitem{MP4} Chin, A. W., Huelga, S. F. \& Plenio, M. B.
Quantum metrology in non-Markovian environments.
\emph{Phys. Rev. Lett.} \textbf{109}, 233601 (2012).

\bibitem{pe1} Dorner, U., Demkowicz-Dobrza\'{n}ski, R., Smith, B. J., Lundeen, J. S., Wasilewski, W., Banaszek, K. \& Walmsley, I. A.
Optimal quantum phase estimation.
\emph{Phys. Rev. Lett.} \textbf{102}, 040403 (2009).

\bibitem{pe2} Kacprowicz, M., Demkowicz-Dobrza\'{n}ski, R., Wasilewski, W. Banaszek, K. \& Walmsley, I. A.
Experimental quantum-enhanced estimation of a lossy phase shift.
\emph{Nature Photon.} \textbf{4}, 357 (2010).

\bibitem{prl2} Taddei, M. M., Escher, B. M., Davidovich, L. \& de Matos Filho, R. L.
Quantum speed limit for physical processes.
\emph{Phys. Rev. Lett.} {\bf110}, 050402 (2013).

\bibitem{MP2} del Campo, A., Egusquiza, I. L., Plenio, M. B. \& Huelga, S. F.
Quantum speed limits in open system dynamics.
\emph{Phys. Rev. Lett.} \textbf{110}, 050403 (2013).



\bibitem{SLD} Braunstein, S. L. \&  Caves, C. M.
Statistical distance and the geometry of quantum states.
\emph{Phys. Rev. Lett.} \textbf{72}, 3439 (1994).

\bibitem{seth} Giovannetti, V., Lloyd, S. \& Maccone, L.
Advances in quantum metrology.
\emph{Nature Photon.} {\bf5}, 222 (2011).

\bibitem{anti} Facchi, P., Nakazato, H.  \& Pascazio, S.
From the quantum Zeno to the inverse quantum Zeno effect.
\emph{Phys. Rev. Lett.} \textbf{86}, 2699 (2001).



\end{thebibliography}
\end{document}